\documentclass[
superscriptaddress,
twocolumn,
 amsmath,amssymb,
 aps,
prstab,
floatfix,
longbibliography
]{revtex4-1}

\usepackage{graphicx}
\usepackage{dcolumn}
\usepackage{bm}
\usepackage{xcolor}
\usepackage{hhline}
\usepackage{siunitx}
\usepackage[normalem]{ulem}
\usepackage[colorlinks]{hyperref}
\hypersetup{
    colorlinks=true,
    linkcolor=blue,
    citecolor=blue,
    filecolor=blue,      
    urlcolor=blue,
}
\usepackage{amsmath}
\usepackage{multirow}
\usepackage{comment}
\usepackage{subcaption}
\usepackage{soul}
\usepackage{xcolor}

\newcommand{\Vt}[1]{$V_0/V_t=\SI{#1}{\percent}$}

\begin{document}

\title{Linear voltage recovery after a breakdown in a pulsed dc system}



\author{Anton Saressalo}
\email{anton.saressalo@helsinki.fi}
\affiliation {Helsinki Institute of Physics and Department of Physics, University of Helsinki,
PO Box 43 (Pietari Kalmin katu 2), 00014 Helsingin yliopisto, Finland}

\author{Dan Wang}
\affiliation {Helsinki Institute of Physics and Department of Physics, University of Helsinki,
PO Box 43 (Pietari Kalmin katu 2), 00014 Helsingin yliopisto, Finland}
\affiliation{State Key Laboratory of Electrical Insulation and Power Equipment, Xi’an Jiaotong University, Xi’an, China}

\author{Flyura Djurabekova}%
\affiliation {Helsinki Institute of Physics and Department of Physics, University of Helsinki,
PO Box 43 (Pietari Kalmin katu 2), 00014 Helsingin yliopisto, Finland}

\date{\today}
\begin{abstract}
\noindent

Breakdowns may occur in high-voltage applications even in ultrahigh vacuum conditions. Previously, we showed that it is important to pay attention to the post-breakdown voltage recovery in order to limit the appearance of secondary breakdowns associated with the primary ones. This can improve the overall efficiency of the high-voltage device. In this study, we focus on the optimization of the linear post-breakdown voltage recovery, with the principle aim of alleviating the problem of the secondary breakdowns. We investigate voltage recovery scenarios with different starting voltages and slopes of linear voltage increase by using a pulsed dc system. We find that a higher number of pulses during the voltage recovery produces fewer secondary BDs and a lower overall breakdown rate. Lowering the number of pulses led to more dramatic voltage recovery resulting in higher breakdown rates. A steeper voltage increase rate lead to a more localized occurrence of the secondary breakdowns near the end of the voltage recovery period. It was also found that the peak BD probability is regularly observed around \SI{1}{\second} after the end of the ramping period and that its value decreases exponentially with the amount of energy put into the system during the ramping. The value also decays exponentially with a half-life of \SI{1.4+-0.3}{\milli\second} if the voltage only increased between the voltage recovery steps.
\end{abstract}

\maketitle

\section{Introduction}
Vacuum arc breakdown (BD) is an event where a conductive plasma channel forms between two metal surfaces separated by a vacuum gap~\cite{Boxman1995HandbookApplications}. Such events are common in any application utilizing high electric fields, including particle accelerators~\cite{Aicheler2018CLICPlan,Pesce2011PassiveExperiment}, radio-frequency power sources~\cite{Catalan-Lasheras2014ExperienceCERN,Cahill2018HighCavities} and vacuum interrupters~\cite{Schade2003NumericalField}. 

The frequency of BD events affects the efficiency of these machines and limits their performance. Therefore, pivotal in minimizing the BD frequency is not only the understanding of the basic mechanisms driving the system towards the formation of a BD spot, but also the BD characteristics and their possible correlation with macroscopically adjustable parameters. Such understanding contributes to the development of safe and cost-efficient high voltage applications. Although the BD phenomenon has been studied for more than a hundred years~\cite{Hull1928SomeVacuum,Snoddy1931VacuumDischarge,Hoffman1997HandbookTechnology,Aicheler2012AReport}, the details of the process and its different stages are still extensively debated~\cite{Antoine2011ElectromigrationAccelerators,Latham1995HighPractice,Engelberg2018StochasticFields,Anders2009CathodicCondensation}

One of the plausible scenarios of development of a BD spot was associated with the activation of dislocations in the presurface region under applied high electric fields~\cite{Pohjonen2013DislocationField,Nordlund2012DefectFields,Engelberg2018StochasticFields,Jacewicz2020Temperature-DependentCryosystem}. In this approach, the dislocations are proposed to induce irregularities on the flat surface. These irregularities can be further sharpened by the locally enhanced electric fields~\cite{Jansson2016Long-termNanotips}, transforming the irregularities into field emitters. These may produce dark current bursts~\cite{Engelberg2020DarkFields,Lachmann2021StatisticalCurrents} that could eventually lead to BDs~\cite{Kyritsakis2018ThermalEmission,Gao2020MolecularNanotips}.

According to the statistical behavior of BD events, the latter were suggested to be divided into two major groups: primary (pBD) and secondary (sBD) BD events~\cite{Wuensch2017StatisticsRegime,Saressalo2020ClassificationSystem,Saressalo2020EffectSurfaces}. The pBDs were both spatially and temporally independent of the preceding events, whereas the sBDs always follow the pBDs, typically correlating also spatially with the location of the previous BD spot. This suggests that it is very likely that there are different mechanisms contributing to generation of a BD spot of each type. 

We showed recently~\cite{Saressalo2020EffectSurfaces} that the post-breakdown voltage recovery must be performed slowly by a gradual increase in the voltage from some initial value ($V_0$) to the target one ($V_t$), which was crucial for the mitigation of sBD. A sBD was almost guaranteed to appeared if the surface was exposed to the target electric field immediately after the breakdown. We also showed that it was essentially equally optimal to recover from a BD by the voltage increase (voltage ramp) either linearly or in twenty voltage increment steps. Although both scenarios were found to be the best out of the tested ones, they still showed a significant number of sBDs and it is desirable to explore the ways of further improvement of the post-breakdown voltage recovery. 

A prominent feature in the step-wise recovery was seen to be an increased BD probability at the beginning of each ramping step. We have previously identified two main factors responsible for this increase \cite{Saressalo2020EffectSurfaces}: 
the \SI{20}{\second} idle time between the voltage steps and the increase in voltage before the next ramping step. The former was attributed to the redeposition of vacuum residual gases on the surface during the idle time, since the reactivity of the surface increases due to the surface cleaning during conditioning. In our previous experiments, it was not possible to avoid pauses during the voltage recovery procedures as it was required for adjusting to new values.

In this work, we developed our setup to exclude the necessity of pauses and focus on linear voltage ramp scenarios only, since these ramping scenarios were the most promising due to time-efficiency and technical accessibility. Moreover, this new approach allows the separation of the effect of the vacuum residual gases from the surface effects, as it allows studying of BD generation at different voltage increase rates while keeping the vacuum level constant. This allows us to focus on the processes which are triggered by the increased applied electric field. In this study, we focus on the short-term conditioning of the electrode surfaces, which is necessary for reducing the amount of sBDs.

\section{Experimental setup} \label{sec:method}

The voltage recovery characteristics were explored using a Pulsed DC system intended for BD studies. The system is described in detail in Refs.~\cite{Saressalo2020EffectSurfaces,Saressalo2020ClassificationSystem,Profatilova2020BreakdownSystem}. It includes a power supply coupled with a Marx generator which magnifies the input DC voltage and converts it into short square voltage pulses. Repetition rate of \SI{2000}{\hertz} and pulse length of \SI{1}{\micro\second} are used in typical BD experiments. In the system, two cylindrical Cu electrodes are enclosed in a vacuum chamber, separated by a gap of $d=\SI{40}{\micro\meter}$. The top electrode (anode) has a contact area diameter of \SI{40}{\milli\meter} and is connected to the Marx generator via a high-voltage (HV) cable, while the bottom electrode (cathode) has a contact area diameter of \SI{60}{\milli\meter} and is grounded. Together, the electrodes form a parallel plate capacitor with a capacitance of around \SI{700}{\pico\farad} (including the capacitance of the HV cable). Voltages of up to $V=\SI{6000}{\volt}$ ($E=\SI{150}{\mega\volt/\meter}$, assuming $E=V/d$ as argued by Ref.~\cite{Korsback2020VacuumSystem}) can be used in the system. The vacuum pressure was around \SI{7e-8}{\milli\bar} in all of the measurements.

Different voltage recovery scenarios were studied in three sets of measurements, all described in Table~\ref{tab:ramptable}. Each measurement consists of 1000 BDs produced using a constant target voltage. The pulsing voltage remained constant at $V_t$, throughout each measurement set, except for the voltage recovery after each BD. In case a BD occurred during the voltage recovery, the process was restarted from $V_0$. Before each measurement, a short re-conditioning was performed to ensure that the electrode surfaces are in an identical state in the beginning of each of the measurements. Additionally, the second measurement set was performed in a random order to identify possible conditioning effects between the measurements. However, the electrode surface state evolved in between the measurement sets, so the results are only fully comparable within each measurement set, but not across the sets.

\begin{table}[htb]
\caption{\label{tab:ramptable} The voltage recovery scenarios applied in the experiments with the linear voltage ramp (one slope). The first column shows symbol each measurement is denoted as. The total number of ramping pulses $N_{\text{pulses}}$ and the fraction of the starting voltage $V_0$ to the target voltage $V_t$ are shown in the second and third column. The final two columns contain the corresponding overall ramping slopes and the ramp rates (voltage increase per pulse), respectively. The last measurement set included scenarios with pulsing at a constant voltage before the start of the voltage ramp. For these runs $N_{\text{pulses}}$ in each phase are denoted in the second column.}

\begin{ruledtabular}
\centering
\begin{tabular}{ccccc}
\textbf{\#} & \textbf{$N_{\text{pulses}}$}  & $V_0/V_t $ &  \textbf{Voltage slope} & \textbf{\textbf{Increase per pulse}} \\
\hline \colrule
1 & 500      &  \SI{20}{\percent}    & \SI{15360}{\volt/\second} & \SI{8}{\volt/pulse}\\
2 & 1000      &  \SI{20}{\percent}    & \SI{7680}{\volt/\second} & \SI{4}{\volt/pulse}\\
3 & 2000      &  \SI{20}{\percent}    & \SI{3840}{\volt/\second} & \SI{2}{\volt/pulse}\\
4 & 4000      &  \SI{20}{\percent}    & \SI{1920}{\volt/\second} & \SI{1}{\volt/pulse}\\
5 & 8000      &  \SI{20}{\percent}    & \SI{960}{\volt/\second} & \SI{0.5}{\volt/pulse}\\
\hline
I & 2000      &  \SI{80}{\percent}    & \SI{880}{\volt/\second} & \SI{0.4}{\volt/pulse}\\
II & 4000      &  \SI{60}{\percent}    & \SI{880}{\volt/\second} & \SI{0.4}{\volt/pulse}\\
III & 6000      &  \SI{40}{\percent}    & \SI{880}{\volt/\second} & \SI{0.4}{\volt/pulse}\\
IV & 8000      &  \SI{20}{\percent}    & \SI{880}{\volt/\second} & \SI{0.4}{\volt/pulse}\\
V & 10 000      &  \SI{0}{\percent}    & \SI{880}{\volt/\second} & \SI{0.4}{\volt/pulse}\\
\hline
A & 6000+4000      &  \SI{60}{\percent}    & \SI{920}{\volt/\second} & \SI{0.5}{\volt/pulse}\\
B & 4000+6000      &  \SI{40}{\percent}    & \SI{920}{\volt/\second} & \SI{0.5}{\volt/pulse}\\
C & 6000+4000      &  \SI{40}{\percent}    & \SI{1380}{\volt/\second} & \SI{0.7}{\volt/pulse}\\
D & 8000+2000      &  \SI{40}{\percent}    & \SI{2760}{\volt/\second} & \SI{1.4}{\volt/pulse}\\
\end{tabular}
\end{ruledtabular}
\end{table}

\begin{figure}[!htb]
\centering
    \includegraphics[width=\linewidth]{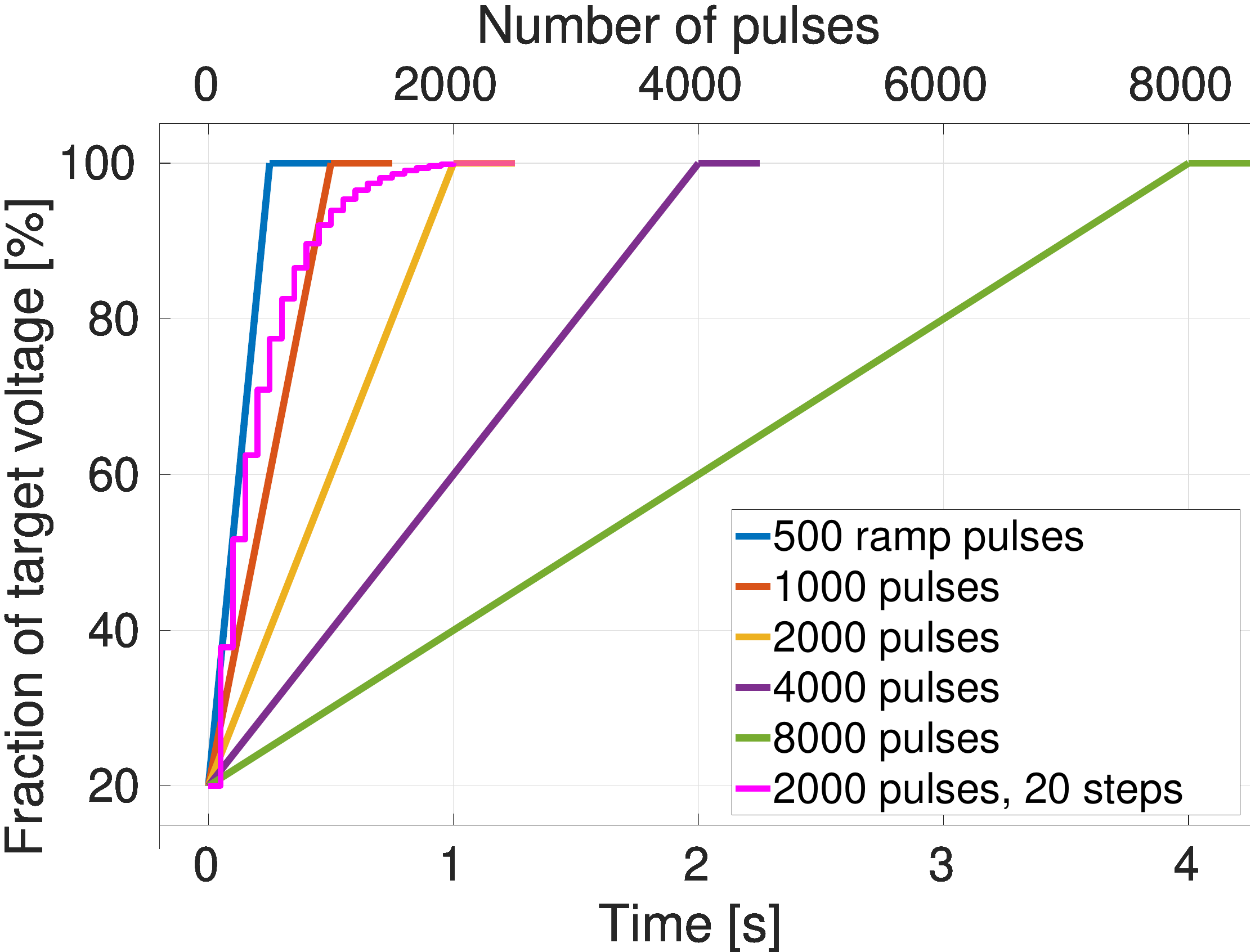}
    \caption{Theoretical voltage increase in each of the ramping scenario in the first set of measurements as a function of time and the number of pulses. With the repetition rate of \SI{2000}{\hertz}, one second equals 2000 pulses. The figure also contains a 20 step ramping scenario used in~\cite{Saressalo2020ClassificationSystem,Profatilova2019BehaviourSystem} for comparison.}
  \label{fig:ramps_ideal}
\end{figure}

In the first set of measurements (scenarios 1--5), ramping scenarios with different voltage slopes were studied. In each measurement, $V_t$ was \SI{4800}{V} (\SI{120}{\mega\volt/\meter}) and the ramping started from \SI{20}{\percent} of the target voltage. Voltage slopes between \SI{1}{\kilo\volt/\second} and \SI{15}{\kilo\volt/\second} were used. The theoretical voltage increase curves are presented in Fig.~\ref{fig:ramps_ideal}.

In the scenarios, the voltage was set to the low value $V_0$ after a BD. For the linear voltage recovery with 2000 pulses, this means that the voltage increases from $V_0=\SI{960}{\volt}$ (\SI{20}{\percent}) to $V_t=\SI{4800}{\volt}$ (\SI{100}{\percent}) within one second (the repetition rate is \SI{2000}{\hertz}). The voltage slope is thus \SI{3840}{\volt/\second}. This is rather fast rate and the system can not adjust to the new value instantaneously, causing delays in the voltage change due to the RC coupling between the power supply and the rest of the circuit. At this rate, the voltage changes from pulse to pulse by $\Delta V = \SI{2}{\volt}$. Although the voltage changes linearly including the time within the \SI{1}{\micro\second} pulse, this change is insignificant, and is only \SI{4}{\milli\volt}. Hence, the change from pulse to pulse is more drastic, which causes the deflection in the shape of the voltage ramp from the linear one.

The second set of measurements (I--V in Table~\ref{tab:ramptable}) focused on the effect of the starting voltage. The slope of the voltage increase remained the same, but the starting voltages ranged from \SI{0}{\percent} to \SI{80}{\percent} of $V_t$ (\SI{4400}{\volt}). Note that scenario IV ($N_{\text{pulses}}=8000$ and \Vt{20}) has identical parameters to scenario 5, except for the slightly different $V_t$. In the last set of measurements (denoted  A--D), the voltage ramp was combined with pulsing at a constant non-zero voltage prior to the voltage increase. Each measurement had the same amount of pulses. $V_t$ of \SI{4600}{\volt} was used in this set of measurements.

Additionally, the shape of the actual voltage increase was studied and compared to the theoretical one via bypassing the Marx generator and measuring directly the output voltage of the power supply across a high-power capacitor of similar capacitance compared to that of the electrode pair. Using a capacitor instead of the Cu electrode pair ensures that no BDs occur during the measurement. Bypassing the Marx generator assumes that the shape of the voltage increase curve is fully determined by the limitations of the power supply and its RC coupling, as the generator is only used to magnify the incoming voltage by a constant factor.

\section{Results}

\begin{figure}[!htb]
\centering
 \includegraphics[width=\linewidth]{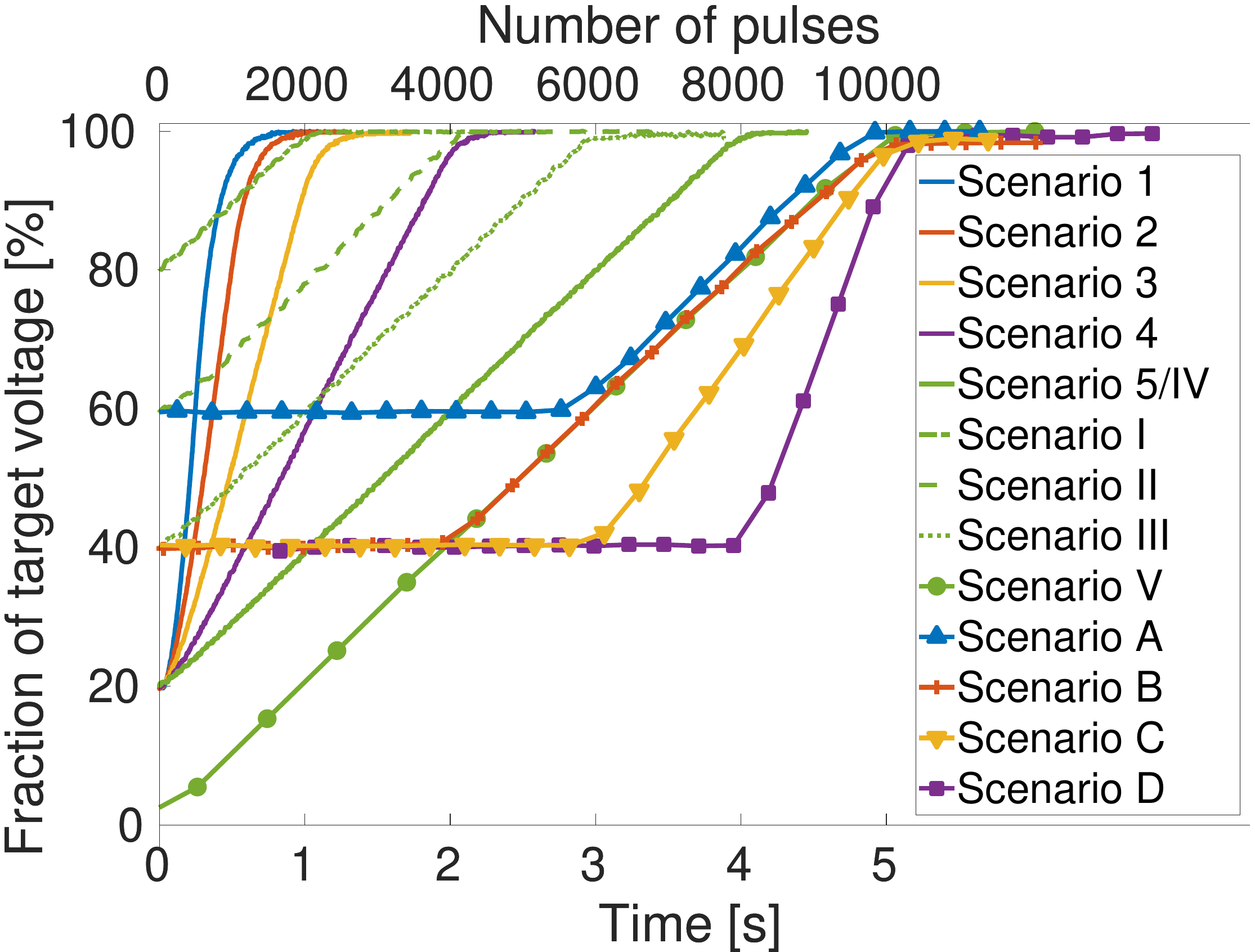}
  \caption{Actual measured voltage increase in each of the ramping scenario as a function of time and the number of pulses. The measurement was performed by using the power supply to increase the voltage to \SI{1000}{\volt} over a circuit with a capacitance of \SI{650}{\pico\farad}.}
  \label{fig:ramps_measured}
\end{figure}

The measured voltage recovery curves for each of the ramping scenarios are shown in Fig.~\ref{fig:ramps_measured}. In each measurement, the target voltage was \SI{1000}{\volt} and the capacitance \SI{650}{\pico\farad}.

In the figure, we see that the actual voltage ramping follows the curves that are slightly different from the theoretical ones of \ref{fig:ramps_ideal} as it was explained in Sec.~\ref{sec:method}. The steeper the ramping, the more dramatic the delay caused by the RC coupling is. It also affects the time, when the voltage reaches $V_t$. For instance, in the scenario with 500 pulses, the actual voltage in the system is only at \SI{60}{\percent} of $V_t$ at the time when the ramping should theoretically end (\SI{0.25}{\second)}. For the other ramping scenarios of the first set, with 1000, 2000, 4000, and 8000 pulses, these values are \SI{80}{\percent}, \SI{91}{\percent}, \SI{97}{\percent} and \SI{99}{\percent} of V$_t$, respectively. For the step-wise voltage ramping scenario with 20 steps, the theoretical and measured curves coincide, since a short pause was used in these experiments in between the voltage increase and the start of pulsing to properly adjust the power supply to the new voltage.

\begin{figure*}[!htb]
\centering
    \includegraphics[width=\textwidth]{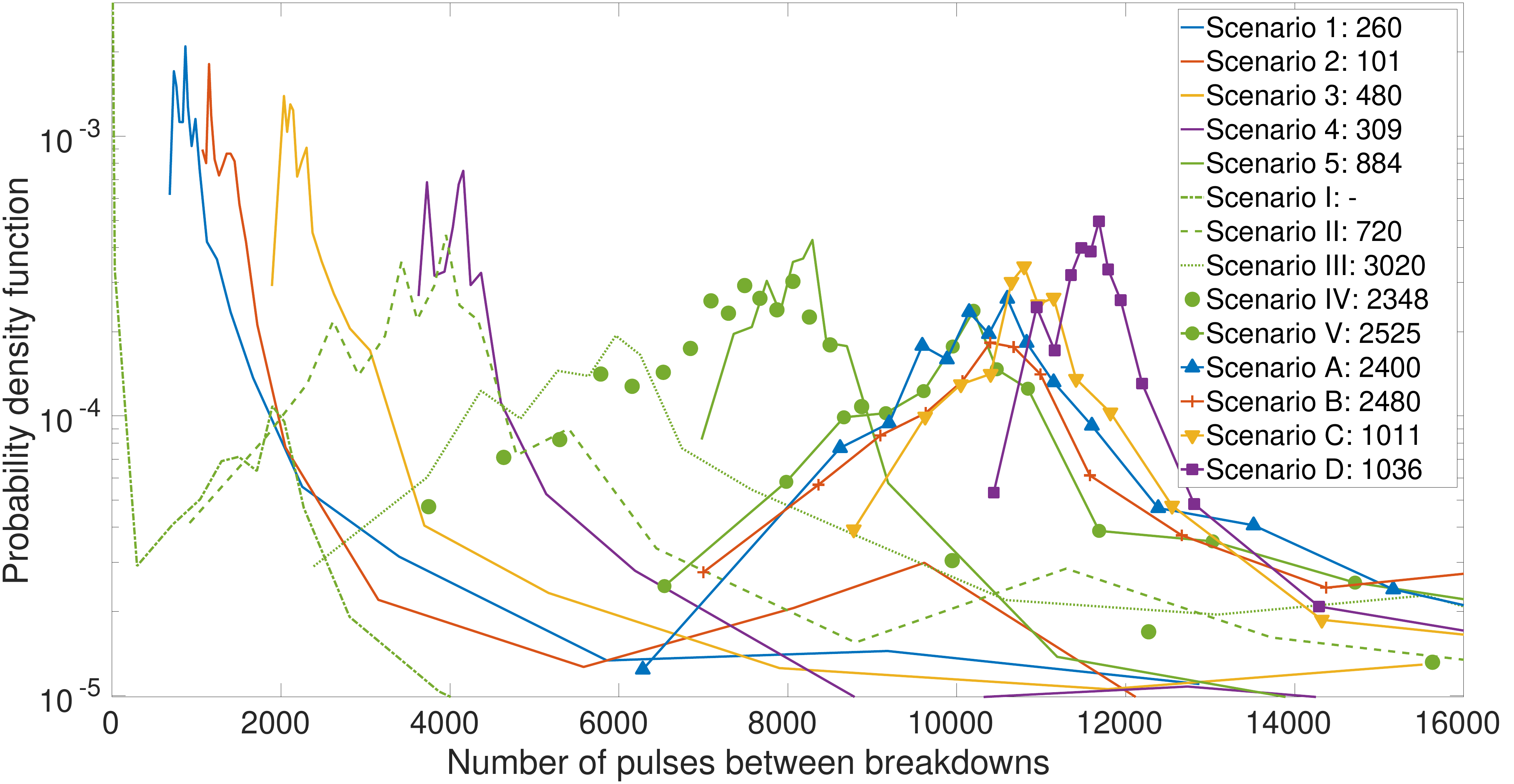}
    \caption{Probability density function for a BD to occur at a number of pulses after the previous BD event for the different voltage recovery scenarios. The number in the legend after each scenario shows the full width at half maximum value for the peak (in pulses).}
  \label{fig:PDF_combined}
\end{figure*}

To determine the best voltage recovery scenario, we analyze the BD rate measured as the number of BDs per pulse (BDR), the percentage of sBDs out of all BDs ($N_{\text{sBD}}$), the mean number of consecutive BDs between two pBDs (including the pBD) ($\mu_{\text{sBD}}$) and $\mu'_{\text{sBD}}$, which is defined as the fraction $N_{\text{sBD}}/N_{\text{pBD}}$, excluding those pBDs that were followed by no sBDs. The results are shown in Table~\ref{table:results}. The uncertainties were estimated by the standard error of the mean. The sBDs were determined by fitting a two-term exponential curve to the probability density function (PDF) of the number of the pulses between two consecutive BD events (Fig.~\ref{fig:PDF_combined}). As introduced in Refs.~\cite{Wuensch2017StatisticsRegime,Saressalo2020EffectSurfaces}, the cross-point $N_{\text{cross}}$ of the two exponential curves was used to classify the events so that the BDs that occurred with the smaller number of pulses in-between ($N < N_{\text{cross}}$) were determined as sBDs, while the pBDs happen with longer waiting time, i.e., more pulses, in between the two BD events. The exponential coefficients $\alpha$ and $\beta$ correspond to the breakdown rates of pBDs and sBDs, respectively, and were found from the fit to the PDFs in Fig. \ref{fig:PDF_combined} as 
\begin{equation}
\text{PDF} = A \exp{(-\alpha N)} +B \exp{(-\beta N)}   
\end{equation}
\cite{Wuensch2017StatisticsRegime}, where $N$ is the number of pulses between consecutive BDs. It must be noted that the results are fully comparable only within each set of measurements, as the surface state of the electrodes may have changed in between the measurement sets.

\begin{table*}[htb]
\begin{ruledtabular}
\centering
\caption{\label{table:results} Key figures for the BD experiments with different ramping scenarios. The first column denotes the measurement id and the second one shows the order the measurements were performed in. Values of breakdown rate (BDR), two-exponential PDF fit parameters ($\alpha$ and $\beta$), cross-point of the two exponentials ($N_{\text{cross}}$), percentage of sBDs ($N_{\text{sBD}}$), the average number of BDs between two pBDs (including the previous pBD) ($\mu_{\text{sBD}}$) and the fraction of the number of sBDs to pBDs for those series with at least one sBD ($\mu'_{\text{sBD}}$) for each scenario are listed in the table. The two-exponential fit could not be performed for scenario I, so the values denoted with $^\dagger$ are not fully comparable to the other measurements.}
\begin{tabular}{c c c c c c c c c c c c}
\# & Order & Scenario & $V_0/V_t$ [\%]   & BDR [bpp] & $\alpha$ & $\beta$ & N$_{\text{cross}}$    & N$_{\text{sBD}}$ [\%]   & $\mu_{\text{sBD}}$ & $\mu'_{\text{sBD}}$\\ 
\hline \colrule
1 & 1 & 500 pulses      & 20 & \num{6.46e-6} & \num{9.4e-5}           & \num{0.003}           & \num{2067}           & \num{30.2}   & \num{2.06+-0.10}  & \num{3.6+-0.7}            \\
2 & 2 & 1000 pulses     & 20 & \num{3.59e-6} & \num{4.2e-5}           & \num{0.003}           & \num{2607}           & \num{26.0}   & \num{1.91+-0.10}  & \num{2.7+-0.7}            \\
3 & 3 & 2000 pulses     & 20 & \num{4.19e-6} & \num{7.4e-5}           & \num{0.002}           & \num{3686}           & \num{23.1}   & \num{1.76+-0.10}  & \num{3.1+-1.0}            \\
4 & 4 & 4000 pulses     & 20 & \num{3.48e-6} & \num{1.8e-5}           & \num{0.003}           & \num{5588}           & \num{22.6}   & \num{1.53+-0.05}  & \num{2.2+-0.5}            \\
5 & 5 & 8000 pulses     & 20 & \num{2.02e-6} & \num{1.7e-5}           & \num{0.002}          & \num{10489}          & \num{21.4}    & \num{1.35+-0.03}  & \num{1.8+-0.4} \\
\hline
I$^\dagger$ & 7 & 2000 pulses     &   $80$     & \num{7.34e-5} & -           & -          & -          & \num{77.4}$^\dagger$    & \num{8.0+-0.7}$^\dagger$  & \num{12+-4}$^\dagger$ \\
II & 9 & 4000 pulses     &   $60$     & \num{3.60e-5} & \num{5.5e-5}           & \num{0.002}           & \num{5378}           & \num{58.7}   & \num{4.3+-0.4}  & \num{7+-3}            \\
III & 10 & 6000 pulses     &   $40$      & \num{3.75e-5} & \num{4.2e-5}           & \num{0.002}           & \num{7284}           & \num{39.9}   & \num{2.5+-0.2}  & \num{4.1+-1.4}            \\
IV & 6 & 8000 pulses     &   $20$      & \num{2.81e-5} & \num{3.7e-5}           & \num{0.001}           & \num{10352}           & \num{62.3}   & \num{3.7+-0.2}  & \num{4.7+-0.6}            \\
V & 8 & 10 000 pulses   &   $0$      & \num{3.30e-5} & \num{4.8e-5}           & \num{0.001}           & 
\num{11943}           & \num{45.4}   & \num{2.5+-0.1}  & \num{3.4+-0.7}            \\
\hline
A & 11 & 6000 pulses flat + 4000 pulses ramp  &   $60$     & \num{5.48e-5} & \num{6.2e-5}           & \num{0.001}          & \num{12755}          & \num{70.2}    & \num{7.5+-1.0}  & \num{13+-7} \\
B & 32 & 4000 pulses flat + 6000 pulses ramp     &   $40$     & \num{3.31e-5} & \num{4.1e-5}           & \num{0.001}           & \num{12228}           & \num{42.8}   & \num{2.6+-0.3}  & \num{5+-2}            \\
C & 13 & 6000 pulses flat + 4000 pulses  ramp     &   $40$      & \num{1.39e-5} & \num{2.8e-5}           & \num{0.001}           & \num{13404}           & \num{39.6}   & \num{1.84+-0.08}  & \num{2.9+-0.6}            \\
D & 14 & 8000 pulses flat + 2000 pulses ramp    &   $40$      & \num{8.84e-6} & \num{1.0e-5}           & \num{0.001}           & \num{14739}           & \num{38.4}   & \num{1.96+-0.13}  & \num{2.8+-0.9}            \\
\end{tabular}
\end{ruledtabular}
\end{table*}

Fig.~\ref{fig:PDF_combined} shows the PDFs as functions of the number of pulses between two consecutive BDs. We see that the location of the BD peak probability correlates with the number of the ramping pulses and is typically registered during the early pulses at the target voltage. We notice not only a shift in the peak location towards the greater number of pulses between consecutive BDs, but also the change of the shape of the peak probability. The more gradually the ramping is performed, the lower the probability of a BD to happen at the end of the ramping process. Hence the concentration of the BD around the end of the ramping stage is lower. This can be observed in the numbers shown in legend that are the full width at half maximum values for the PDF peaks (in pulses). The broader this value, the more evenly the BDs are distributed over the different numbers of pulses between the consecutive BDs. This correlation is particularly obvious in the first set of the measurements 1--5, where the effect of ramping slope was analyzed.

Comparing the results for the scenarios 1--5 presented in Table~\ref{table:results} and Fig.~\ref{fig:PDF_combined}, we note that the lowest values in BDR, $N_{\text{sBD}}$ and $\mu_{\text{sBD}}$ were obtained for the scenarios with the most pulses in the voltage recovery. However, the exponential coefficients $\alpha$ and $\beta$ were relatively similar for all scenarios. This consistency indicates the same nature of the processes responsible for the sBD and the pBD events. The striking consistency in the values of the $\beta$ coefficient, (the rate of sBDs) shows that this process does not depend on the particular ramping scenario due to the correlation existing between the subsequent events. Slight gradual decrease of the $\alpha$ coefficient with increase of the number of ramping pulses indicates a quieter conditioning process via the pBDs only. We also note that despite the similarities in $\alpha$ and $\beta$ coefficients for all the measurements in this series, both values are consistently lower for the scenario with 8000 voltage recovery pulses.

The similar $\alpha$ and $\beta$ values in all the measurements lead to the consistent shift of about 2000 pulses (\SI{1}{\second}) in the $N_{\text{cross}}$ towards a greater number of pulses after the voltage recovery is finished. The combination of the smallest $\alpha$ and $\beta$ values in the recovery scenarios with the most pulses ($\geq 8000$) resulted also in the largest shift in $N_{\text{cross}}$ towards the greater number of pulses between the consecutive BDs. Despite this, we conclude that in our measurements almost all BDs, separated by more than an extra $\sim$\SI{1}{\second} after the voltage recovery period, are pBDs.

\begin{figure}[!htb]
\centering
    \includegraphics[width=0.49\textwidth]{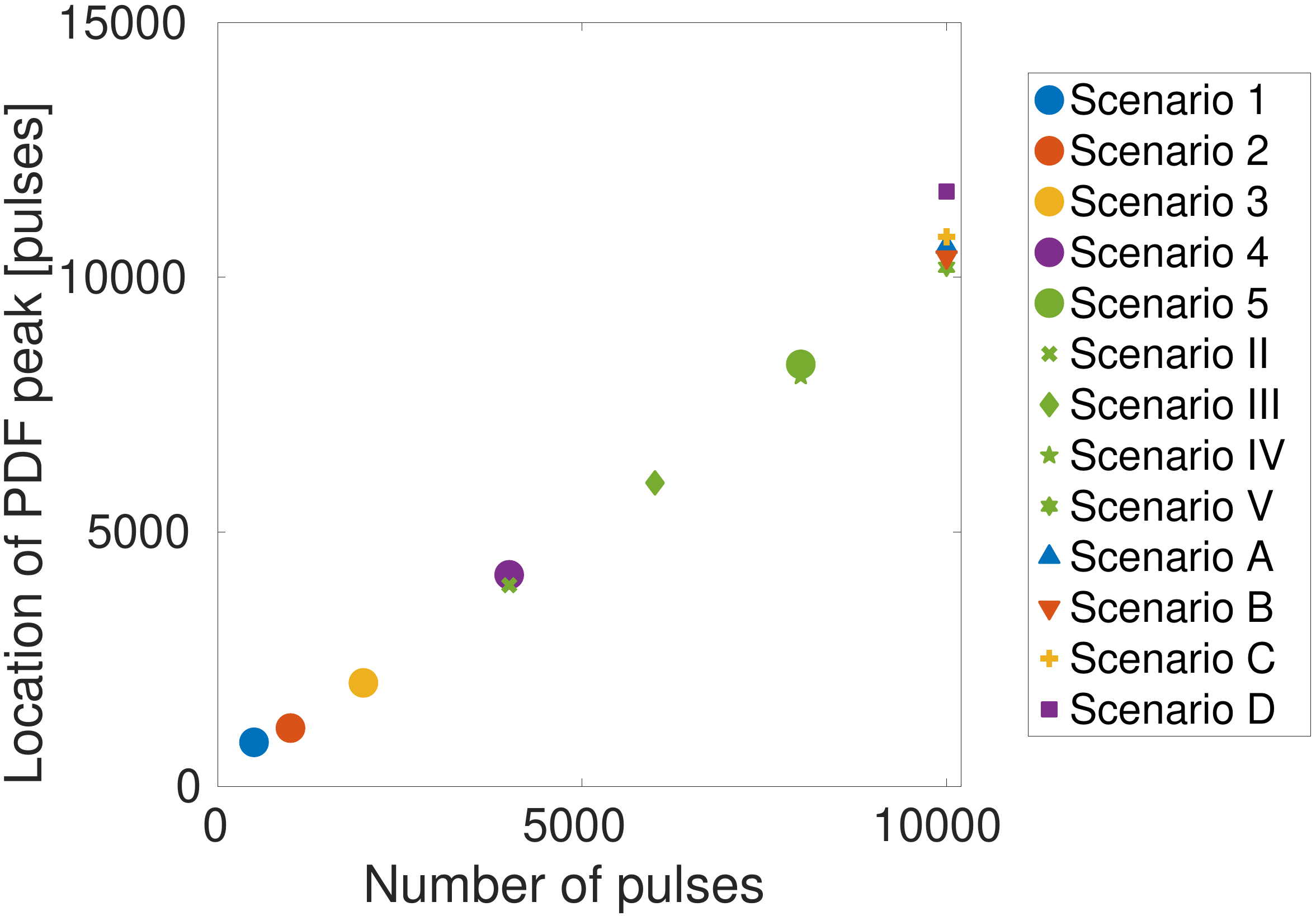}
    \caption{Location of the maximum value of each PDF in~\ref{fig:PDF_combined} as a function of the number of pulses in the voltage recovery scenario.}
  \label{fig:peakPulses}
\end{figure}

In Fig.~\ref{fig:peakPulses}, we see that the location of the maximum value of each PDF correlates strongly with the number of voltage recovery pulses. The maximum BD probability is typically reached a few hundred pulses after the voltage has reached its target value.

Since the results of scenarios 1--5 clearly indicate that the slope of the voltage ramp affects the BD activity during the short-term conditioning, we now turn to the analysis of the results obtained in the measurements performed with the purpose of reducing the total ramping time while keeping the lowest slope. In the next measurement set (scenarios I--V), we start the voltage recovery from different fractions of $V_t$ while keeping the slope of the voltage increase unchanged.
 
The results in Table~\ref{table:results} show that the shortest ramping time in this set of the measurements (see scenario I, \Vt{80}) resulted in by far the worst performance as more than \SI{50}{\percent} of the BDs occurred on the first pulse after the previous BD. This made it impossible to fit the two-exponential curve or to estimate the FWHM for the PDF of scenario I. The scenarios with other starting voltages resulted in the BD behaviors with the similarly low number of sBDs, although the scenario with \Vt{60} resulted in the highest BD rate and the longest sBD series. It is clear that pulsing at the voltage below \Vt{40} was not contributing to the improvement of the conditioning procedure.

The experiments I--V were not performed in the logical order to identify possible conditioning effects. However, the results show no correlation with the measurement order, only with the measurement parameters.

We next aim to verify the effect of the number of pulses during the voltage recovery stage. For this, we performed one more set of experiments (scenarios A--D), where we fix the total number of pulses and use two different combinations: fixed ramping slope and a varied initial voltage value or fixed initial value and a varied ramping slope. In these, the voltage  ramping did not start immediately with the pulsing, but with a delay, i.e. after pulsing at a flat value of \Vt{60} and \Vt{40} for \SI{3}{\second} and at \Vt{40} for \SI{2}{\second} and \SI{4}{\second}. 
The results of these measurements clearly show that long pulsing at \Vt{60} resulted in the worst performance in this series of experiments. We see that the long pulsing at a relatively high voltage value, although with no ramping, increased the unstable conditioning performance of the PDF. Namely, we see that the distribution has the longest left-hand tail before the peak, which means that the BDs start before the voltage is ramped to the target value and continue long after $V_t$ is reached.

Long pulsing at \Vt{40}, however, with different combinations of the delay time and the ramping slope resulted in very similar performance, however, clearly better than the performance of the scenario III, where the recovery scenario was similar, but with no delay in applying the ramping. Even more surprisingly, increasing the delay time (hence increasing the voltage ramp at the end of the voltage recovery procedure) did not result in worsening of the conditioning performance, but on contrary, reduced the BD rate, while keeping the percentage of the sBDs the same low, and shortened the length of the sBD series. 

It is a surprising result, which is rather counter-intuitive, taking into account our previous observation of dependence of the conditioning process on the voltage slope. In Fig.~\ref{fig:ramps_measured}, we clearly see that the scenario D had the steepest voltage increase slope, although it resulted in BDs occurring after the largest number of pulses within the ramping period and practically in no BD before the voltage reached its target value. It also had the narrowest PDF peak (the appearance of BDs is more localized around its peak value) and a fast decay towards the higher number of the pulses between subsequent BDs.

Based on these results, we conclude that the surface is self-cleaning during the pulsing at relatively low voltage values. This means that that even the steep rise of the voltage during the voltage recovery stage results in mainly pBDs, only rarely followed by a cascade of sBDs.

\begin{figure}[!htb]
\centering
    \includegraphics[width=0.49\textwidth]{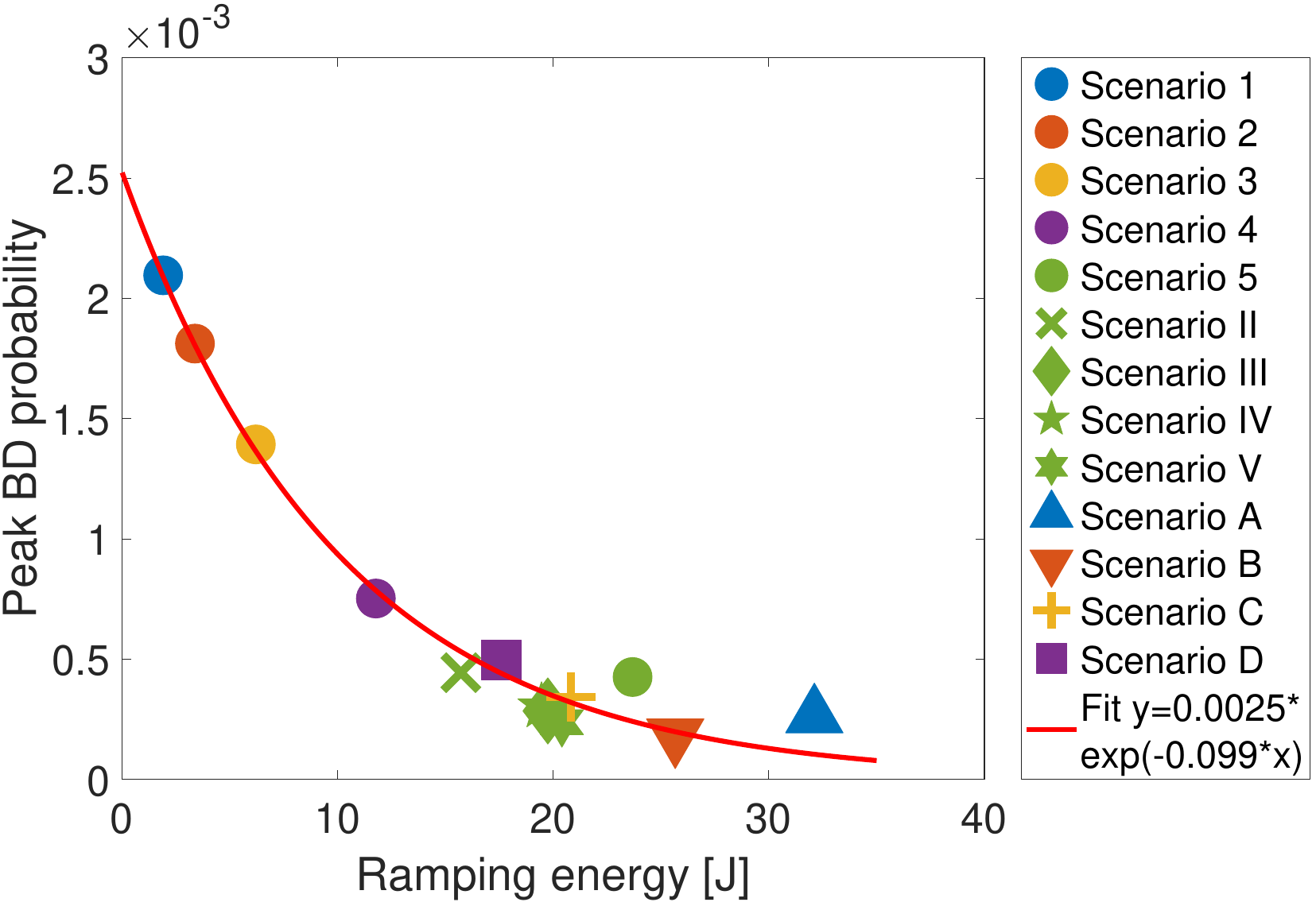}
    \caption{Peak BD probability during linear voltage recovery as a function of the ramping energy. Also a fit describing an exponential decay is shown.}
  \label{fig:peakBDprobs_vs_energy}
\end{figure}

Fig.~\ref{fig:peakBDprobs_vs_energy} shows the peak BD probabilities from Fig.~\ref{fig:PDF_combined} versus the total energy $E$ that is entered into the system via pulsing during the voltage recovery. The energy was calculated by summing the energy $E_i$ of each voltage pulse $i$
\begin{equation}
    E = \sum_{i=1}^{N_{\text{Pulses}}} \frac{1}{2} C V_i^2,
\end{equation}
where $C=\SI{650}{\pico\farad}$ is the capacitance of the system and $V_i$ is the voltage of each ramping pulse. Scenario I is excluded from the figure since its peak BD probability is several orders of magnitude higher compared to the other measurements. It is interesting to observe that all points obtained in different measurements follow closely the exponential dependence fitted as

\begin{equation}
\text{PDF}_{\text{peak,slopes}} = A\exp(BE)
\end{equation}
with the fitting coefficients $A=\num{2.5+-0.2e-3}$ and $B=\num{-0.099+-0.011}$ and the coefficient of determination $R^2 = 0.98$.

\begin{figure}[!htb]
\centering
    \includegraphics[width=0.49\textwidth]{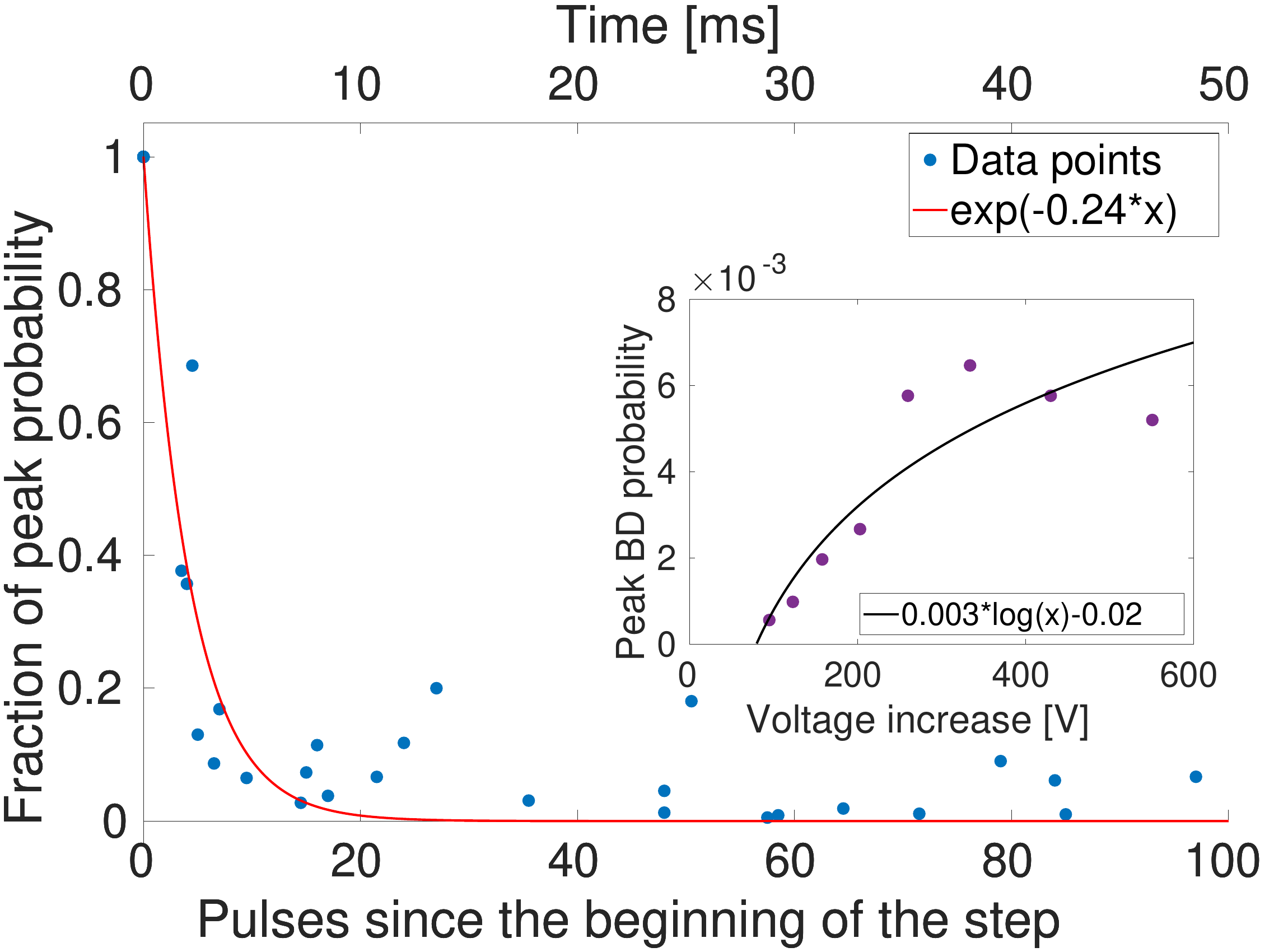}
    \caption{Decay of BD probability as a function of pulses (or time) since the beginning of the ramping step. The red line shows an exponential fit on the data. The inset shows the relationship between the peak BD probability during a ramping step and the increase of voltage since the previous step.}
  \label{fig:ramp_probs}
\end{figure}

Finally, we revisit the step-wise voltage recovery measurement ($V_t$ is reached using 2000 pulses in 20 steps), which is commonly used in metal surface BD conditioning experiments as reported in~\cite{Saressalo2020EffectSurfaces,Profatilova2020BreakdownSystem}. In Fig.~\ref{fig:ramp_probs}, we analyze the data within each step (between two subsequent changes in the voltage values) to determine the decay rate of the BD probability within a ramping step. In the scenario, the step length is 100 pulses, during which the voltage stays constant. In the figure, we plot the PDF produced by a step-wise voltage recovery scenario as a function of the number of pulses between two consecutive BDs. We noticed that, during the ramping process, the PDF shows a saw-tooth behavior, as was observed in Refs.~\cite{Saressalo2020ClassificationSystem,Saressalo2020EffectSurfaces}. The behaviour shows a strong increase in the PDF at the first pulses after the voltage change and a rapid decay until the end of the step. We now analyze the decay rate of the BD probability $P$ by fitting it to an exponential function
\begin{equation}
    P(t) = \exp\left(-\frac{\ln{(2)}}{\tau_{1/2}} t \right),
\end{equation}
where $t$ is the time (or the number of pulses) since the beginning of the ramping step. The half-life $\tau_{1/2}$ was found to be \SI{2.9+-0.6}{pulses}, which equals \SI{1.4+-0.3}{\milli\second} (coefficient of determination $R^2=0.95)$. This shows a very sharp decay in the PDF after the change of the voltage, confirming the strong effect of the idle time during the voltage change. Furthermore, we analyze the relationship between the peak BD probability during a ramping step and the voltage increase after the previous step in the inset of Fig.~\ref{fig:ramp_probs}. This dependence can be fitted to the logarithmic function of the voltage increase between the ramping steps
\begin{equation}
    \text{PDF}_{\text{peak,steps}} = a_{\text{steps}}\log(\Delta V)+b_{\text{steps}}
\end{equation}
are $a_{\text{steps}}=\num{3+-2e-3}$ and $b_{\text{steps}}=\num{-2+-1e-2}$ (coefficient of determination $R^2 = 0.80$).
The non-ideal fitting could be explained by the fact that in this case, not only the voltage increase, but also the idle time between the ramping steps affects the BD probability~\cite{Saressalo2020EffectSurfaces}.

\section{Discussion}
The results shown in Table \ref{table:results} confirm that the overall behavior of most linear ramping scenarios tested in this work are very similar. Since there were no pauses in these experiments, neither during nor after the voltage ramp, the PDF decay slopes seen in Fig.~\ref{fig:PDF_combined} after the peak values are very similar and did not strongly depend on whether a short or long ramping time was used. Quantitatively the shape similarity is confirmed by the similar values of $\alpha$ and $\beta$, i.e., the respective pBD and sBD rates for all the scenarios. However, we noticed that the most quiet and, hence, least detrimental conditioning process is achieved in the longest voltage recovery process, i.e., the one with the lowest ramping slope and the greatest number of recovery pulses. Hence, we separately analyzed which of the two parameters is the most important for optimization of the conditioning process.

Fixing the value of the ramping slope, but aiming to reduce the time of the voltage recovery stage, we performed the measurements with different starting voltages (scenarios I--IV). The results emphasize the importance of the slope, showing very comparable results for almost all the scenarios except for the shortest one. The scenario with $V_0=\SI{80}{\percent}$ performed significantly worse than the others. This can be explained by the fact that starting the voltage recovery at such a high voltage was very close to scenario in which no recovery was used, which was found detrimental in~\cite{Saressalo2020EffectSurfaces}. Surprisingly we did not observe effect of extra pulsing when the voltage was ramped up from zero or from \Vt{20}. This inspired us to verify whether the longer pulsing at relatively high initial voltage value would affect the result.

In the third set of measurements (scenarios A--D), we fixed the total number of pulses of the voltage recovery stage to the largest value of the previous set of experiments (\num{10000}), alternating pulsing at the flat voltage value to the voltage ramping mode. To our surprise, the scenario D with the steepest ramping slope resulted in the least number of sBDs. We believe that in this experiment the combination of long pulsing at relatively high, but still sufficiently low value, added the effect of a gentle cleaning of the surface, which allowed to ramp the voltage even faster than usual without increasing the number of breakdowns. On the contrary, the breakdown rate and the average number of sBDs between the two subsequent pBD decreased dramatically (see Table \ref{table:results}). However, the flat value in the voltage during the initial pulsing plays a role. Similar to scenario I, it was found that a high $V_0$ in scenario A made it perform worse than the others. We also note that in overall, in Fig.~\ref{fig:PDF_combined}, the scenarios A (long pulsing at \Vt{60}) and B (short pulsing at \Vt{40}) produced very similar PDF curves, while the curves of the scenarios C and D (in both, the long pulsing at $V_0$=\SI{40}{\percent}$V_t$) where slightly more localized. 

Although we are not able to compare the results between different sets of experiments quantitatively because of the different state of the used electrodes, we observe a general trend in the shapes and the peak locations of all PDFs. It is clear that, in all the experiments, the location of the peak correlates strongly with the number of pulses used during the voltage recovery stage. Moreover, we notice that the shape of the PDFs in the C and D experiments with the steeper voltage ramping slope is very similar to the shape of the PDFs obtained in the first experiments 1--5, where the slopes of the voltage ramp were also the steepest. The less localized peaks of the I--V and A--B experiments with the very similar slopes confirm our conclusion that the slope of the voltage ramp defines how localized the BD probability is around its maximum value.

Similarity in the PDF shapes of Fig.~\ref{fig:PDF_combined} confirms that the surface conditions were similar during all the experiments, since other parameters of the experiments, such as the gap length and the target voltage $V_t$, remained unchanged. However, we noticed that the BD probability during the early pulses after the voltage reached $V_t$, was increased as the ramping time was shortened. Moreover, these BDs with the smallest number of pulses between them, form their own distribution, which can be seen in Fig.~\ref{fig:PDF_combined} as short tails on the left side of the graph. The shapes and the maxima of these distributions are different for the different ramping slopes of the post-breakdown voltage recovery.

Considering that the right-hand parts of the PDFs in Fig.~\ref{fig:PDF_combined} can be fitted to a Poisson distribution using a two-exponential model, we can assume that both the sBDs and pBDs are produced by distinct Poisson processes. This also implies that each BD event must be independent of the preceding one, i.e., it must appear with equal probability either immediately or after any number of pulses since the previous BD. Existence of the tails in the region of the smallest numbers of pulses between the consecutive BDs, with the peak values PDF$_{\text{peak}}$ indicates of enhancement of the BD probability for these values compared to others, i.e., BDs are more probable to occur right after the end of the ramping process compared to any other pulse number in the region where the voltage is constant $V=V_t$. The difference in PDF$_{\text{peak}}$, consistent with the ramping slopes used during the recovery process, reveals that the surface adjusts differently to the voltage increase depending on the rate of the ramp.

In Fig.~\ref{fig:peakBDprobs_vs_energy}, we found out that the PDF$_{\text{peak}}$, as a function of the energy pumped into the system during the ramping pulses, follows an exponential dependence. The more energy is put into the system during the ramping, the lower the peak BD probability. This suggests that the work performed by the ramping pulses affects the surface and broadens the PDF peak, making the BD occurrence more delocalized in terms of pulses between consecutive BDs. This is why a more localized probability is seen for the shorter voltage recovery scenarios in Fig.~\ref{fig:PDF_combined}. However, when the energy is already significant, the further localization of sBDs during the first pulses slows down.

Furthermore, we also note that the percentage of the sBDs and the average length of the sBD series $\mu_{sBD}$ are also monotonically decreasing with a decrease in the ramping slope. Surprisingly, the average length of the sBD series ($\mu'_{sBD}$), which actually occurred after a pBD, showed much weaker dependence on the ramping slope, fluctuating between 2-3 sBD after a pBD for the scenarios 1--5. It is clear that the shortening of $\mu_{sBD}$ with the decrease of the slope is explained by surface conditioning mainly via less violent and smoother distributed surface pBDs.

As it was mentioned earlier, there is a consistent shift in the cross-point $N_{\text{cross}}$ of the two exponential fits of the PDFs by approximately 2000 pulses (\SI{1}{\second}) after the voltage ramp is finished. However, this shift is stronger for the fast ramp and reduces with the increase of the ramping time. We relate this trend to the delays in the voltage ramping due to RC coupling. As it was shown in Fig.~\ref{fig:ramps_measured},
the target voltage is not reached immediately after the given number of pulses, but with some delay. The delay is greater when the voltage increase is steeper. This suggests that a vast majority of the BDs, that occur after the target voltage is reached and stabilized, are pBDs. The apparent exception of this trend is found in scenario 5, where we see that $N_{\text{cross}}$ in this case is shifted deeper towards the greater number of pulses. However, we draw the attention to the fact that we obtained the shortest series of consecutive sBDs using this voltage recovery scenario, which reduces statistical accuracy of the obtained results. In overall, we found the smaller sBD rates and much lower localization of sBDs at the first pulses after the voltage ramp. However, the $N_{sBD}$ is consistently lower than in any other experiments, which confirms the conclusion about overall conditioning performed via pBDs mainly.

From Fig.~\ref{fig:PDF_combined}, it can be seen that the BDs start occurring closer to the end of the ramping period when the voltage ramping slope is smaller. This indicates that the metal surface can tolerate stronger electric fields if the voltage is ramped slowly. The result shown in Fig.~\ref{fig:peakBDprobs_vs_energy} is also in line with this argument. Here we saw that the BD probability is lower if more energy is put into the system during the early pulses.

As we saw previously in the experiments with the step-wise voltage recovery scenarios, the increase in the voltage value before each following step increased the BD probability drastically during the first few pulses. We showed in Fig.~\ref{fig:ramp_probs} that this decay is very fast with a half-time of \SI{1.4}{\milli\second}. This suggests that any pulse, also in the linear voltage ramp scenarios, with a voltage higher than that of the previous, increases the BD probability for a short time and this increase depends on the difference $\Delta V$ between the pulses. 

The results above can be associated with the existence of short-lived field emitters on the electrode surface. When the surface is exposed to an electric pulse with an amplitude sufficient to initiate the formation of new field-emitting features, fresh emitters can be born emitting electrons into vacuum. Each emitter can either exhaust itself or enter into a runaway process, which can eventually lead to a full BD event~\cite{Kyritsakis2018ThermalEmission}. One of the known factors determining the outcome of the emitter is the electric field strength, which has to be sufficiently high to launch the runaway process~\cite{Kyritsakis2018ThermalEmission}. The hypothesis assumes that the amount of new emitters born is proportional to the change in the electric field strength. During the voltage change, there is a sudden increase in the external electric field strength at the surface, which magnifies the local gradients, driving the atomic migration on the surface. The relationship between the peak BD probability and the ramping energy found in Fig.~\ref{fig:peakBDprobs_vs_energy} also supports this assumption. The work performed by the electric pulses at relatively low voltages can be sufficient to bring out the field emitters, but not enough to start the runaway process. Thus, when the system enters into pulsing with high enough voltages that could cause BDs, many of the possible field emitters are already exhausted, thus lowering the BD probability.

However, more studies would be required to confirm the hypothesis involving the dynamics of the short-living field emitters.

Moreover, it is apparent that in the case of the longest ramping, the system had experienced the largest number of pulses before the voltage reached the target value. Electric fields exert tensile Maxwell stress on metal surfaces~\cite{Griffiths2013IntroductionElectrodynamics}, hence in the pulsing regime, this stress may induce surface vibrations. These vibrations can contribute, e.g. in cleaning of the surface by detachment of the loosely bound impurity atoms, which is another possible way of reducing the BD probability.

The counter-intuitive results between the first and third measurement sets can be explained by the fact that, once the surface is cleaned from the impurity atoms due to the vibrations during the early ramping pulses, a higher strain rate (produced by a higher voltage slope) may actually increase the yield strength of the metal, thus resulting in fewer field emitters that could cause BDs~\cite{Liu2008ACopper,Bandyopadhyay2017Non-equilibriumLoading}. However, this means that the voltage during the early ramping pulses needs to be high enough to produce any cleaning effect, while at the same time the voltage cannot be so high that it would cause BDs. It was found that \SI{40}{\percent} of $V_t$ is a good compromise between these two.

\section{Conclusions}
Different post-breakdown voltage recovery scenarios with linear ramping and no pauses were investigated in this study. The main purpose was to determine the best scenario resulting in the most efficient mitigation of the secondary BDs and to understand the relationship between the rate of the voltage increase and the BD probability.

We observed that the breakdown rate, the fraction of secondary BDs and the average number of secondary BDs per a primary BD in a series were the lowest for the scenarios with the most pulses to reach the target voltage. The smaller the slope of the voltage recovery, the fewer BDs were produced during the recovery.

In all scenarios, the sBD probability was concentrated at the end of the ramping period, regardless of the number of the ramping pulses. The localization of the sBD probability was found to be dependent on the slope of the voltage increase. It was also found that the peak BD probability is dependent on the energy of the ramping pulses: the more energy is inserted into the system during the ramping, the lower the peak BD probability.

The lowest breakdown rate of the fist set (scenarios 1--5) was achieved with the scenario that contains the most gradual voltage recovery and the most voltage recovery pulses. For second set (scenarios I--V), with equal voltage slopes, but variable starting voltages and number of pulses, the three scenarios with 6000, 8000 and \num{10000} pulses and starting at \SIlist{40;20;0}{\percent} of the target voltage, respectively, performed almost equally well.

For the last measurement set (scenarios A--D), the lowest BDR was achieved with a combination voltage recovery scenario with 8000 pulses of pulsing at a flat voltage, \SI{40}{\percent} of the target voltage, followed by a rapid increase to the target voltage over \SI{2000}{pulses}.

\section*{Acknowledgements}
The authors would like to thank the China Scholarship Council (CSC), which supported parts of the work.

\bibliography{refs}
\end{document}